\documentclass[prb,aps,espf,showpacs,twocolumn]{revtex4}
\usepackage{amsmath}
\usepackage{graphicx}
\usepackage{color}

\begin{document}

\title {Strain and correlation induced half-metallic ferromagnetism in orthorhombic  BaFeO$_{3}$}

\author{Gul Rahman}\email{gulrahman@qau.edu.pk}
\author{Saad Sarwar}
\address{Department of Physics,
Quaid-i-Azam University, Islamabad 45320, Pakistan}

\begin{abstract}
Using first-principles calculations, the electronic and magnetic properties of orthorhombic BaFeO$_{3}$ (BFO) are investigated with local spin density approximation (LSDA). The calculations reveal that at the optimized lattice volume  BFO has a lower energy in ferromagnetic state as compared with antiferromagnetic state. At the equilibrium volume, BFO shows metallic behavior, however, under a large tensile strain ($\sim25\%$), BFO shows half-metallic behavior consistent with the integer magnetic moment of $4.0\mu_{\rm{B}}$/fu mainly caused by the $t_{2g}$ and $e_{g}$ electrons of Fe. 
Including a Hubbard-like contribution $U$ (LSDA$+U$) on Fe $d$ states
induced half-metallic bahvior without external strain, which indicates that $U$ can be used to tune the electronic structure of BFO. The magnetic moments remained robust against $\sim 10\%$ compressive and tensile strain. At large compressive (tensile) strain, the half-metallicity of BFO is mainly destroyed by the Fe-$d$ (O-$p$) electrons in agreement with the non-integer value of the magnetic moments of BFO.     
\end{abstract}

\maketitle

Since the discovery of half-metallic materials in 1983 by Groot $et.al$,\cite{groot} many other new materials have been discovered using first-principles calculations because these materials are promising candidates in the area of spin-electronics.\cite{gr2010} 
Perovskite-structure oxides, ABO$_{3}$, are of technological interest due to their numerous and often coupled functionalities, such as 
ferroelectricity, magnetism, multiferroicity, metal-insulator transitions, etc.
The perovskite family is known to have a large variety of properties\cite{2,3,4,5} including spin dependent transport properties \cite{6}. 
Some perovskite materials also show half-metallicity driven by external pressure. 
When the external pressure is below $\sim$ 80\,GPa,$\rm NiCrO_3$ is an antiferromagnetic (AFM) semiconductor and behaves as an AFM half-metal when the external pressure is increased \cite{17}. 
At ambient conditions $\rm Sr_2NiReO_6$ is a semiconductor, and through external pressure, a semiconductor to half-metal transition is observed.\cite{18}	$\rm La_2CoMnO_6$ also transformes to half-metallic state under external pressure\cite{19}.

Fe-containing ferroelectric oxides are also attracting tremendous attention due to their magnetic properties. Particularly, those oxides that exhibit magnetic and ferroelectric characteristics simultaneously can have practical device applications such as spin transistor memories, whose magnetic properties can be tuned by electric fields through the lattice strain effect.\cite{ref1} 
Fe-based perovskite BaFeO$_{3}$ (BFO) normally assumes a hexagonal crystal structure, although various
polymorphs have been observed with oxygen deficiency.\cite{Hombo,Iga,WW}
Recent experiments observed, however, ferromagnetism in cubic BFO, where the metastable cubic structure was shown to be stable down to $8$K.\cite{expt}
Very recently, Li {\it et al.}\cite{LiPRB2012} used DFT$+U$ to investigate the helical structure of cubic ferromagnetic (FM) BFO and found AFM to FM transitions under external pressure.

It has been shown that the
functional behavior in oxides can be enabled or enhanced when they are prepared in the form of thin films. For example, ferroelectricity has been reported in thin films
of SrTiO$_{3}$, which is a quantum paraelectric in its bulk form, and enhanced polarization has been observed in thin films of ferroelectric BaTiO$_{3}$.\cite{bto,sto}
In both cases, the improved behavior is attributed to the bi-axial epitaxial
strain introduced when the film is grown coherently on a substrate with a different lattice constant.
{{As there are no theoretical and experimental work have been performed on orthorhombic BaFeO$_{3}$, and the possibility of orthorhombic phase can not be ignored when we synthesis BaFeO$_{3}$. Such orthorhombic phase is already exists in other perovskites, e.g., BiFeO$_{3}$, BaTiO$_{3}$, etc.} Therefore, we focus on orthorhombic BaFeO$_{3}$ and  show that orthorhombic BaFeO$_{3}$ has a negative formation energy, which indicates that we can not ignore this crystallographic phase of BaFeO$_{3}$. In particular, we also show that that external strain and strong correlation can induce half-metallic ferromagnetism in orthorhombic BaFeO$_{3}$.}

The First-principles calculations based on density functional theory are performed with  plane-wave and pseudopotential method as implemented in Quantum Espresso package.\cite{20} The exchange correlation effects are treated within local spin density approximation(LSDA). Later on, the on-site Coulomb potential $U$(=5.0eV) \cite{21} has also been added in LSDA to perform LSDA+$U$ calculations to correctly describe the electronic structure of orthorhombic BaFeO$_{3}$ (denoted as BFO in the manuscript). The ultrasoft pseudopotentials are used to describe the core-valence interactions. The valence wave functions and the electron density are described by plane-wave basis sets with kinetic energy cutoff of 30 Ry with $12\times12\times12$ Monckhorst-Pack grid. All the computational parameters were carefully converged, {and the convergence criterion for the electronic energy is $10^{-6}$ eV.}

\begin{figure}
\hspace{-1cm}
\includegraphics [width=0.22\textwidth]{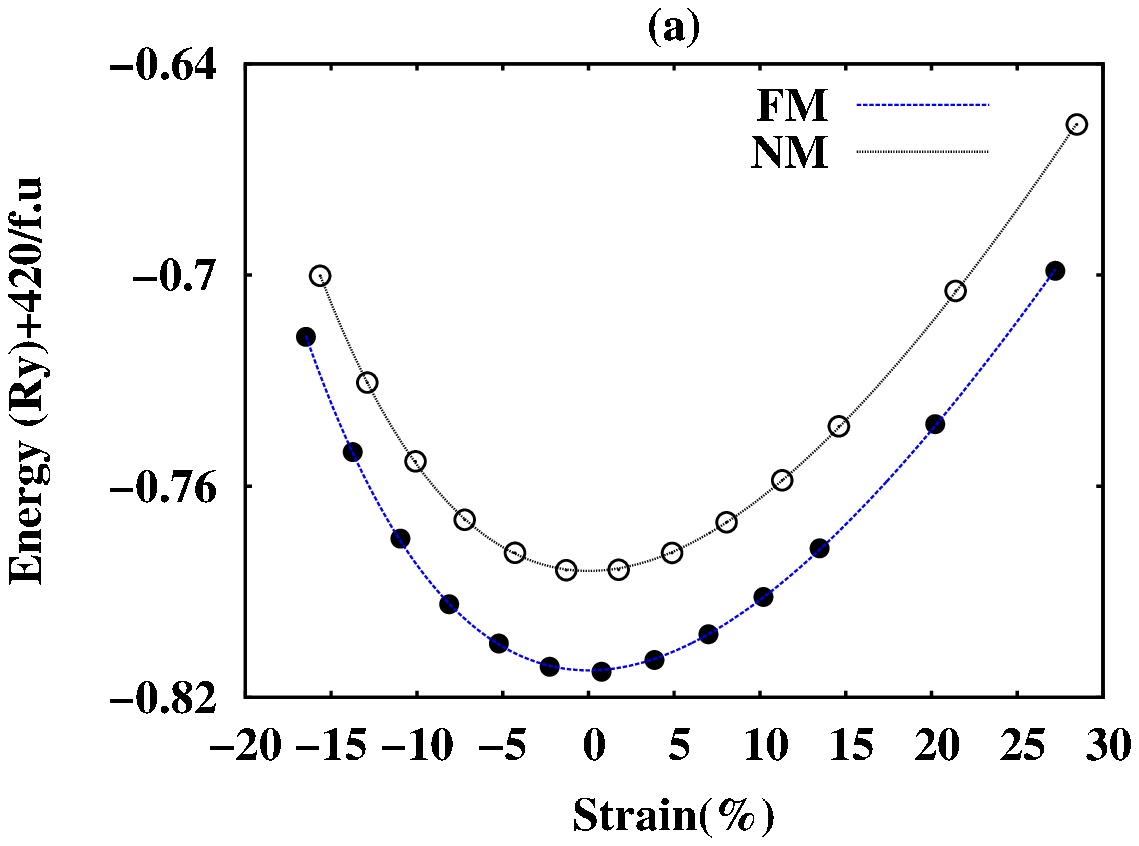}
\includegraphics [width=0.22\textwidth]{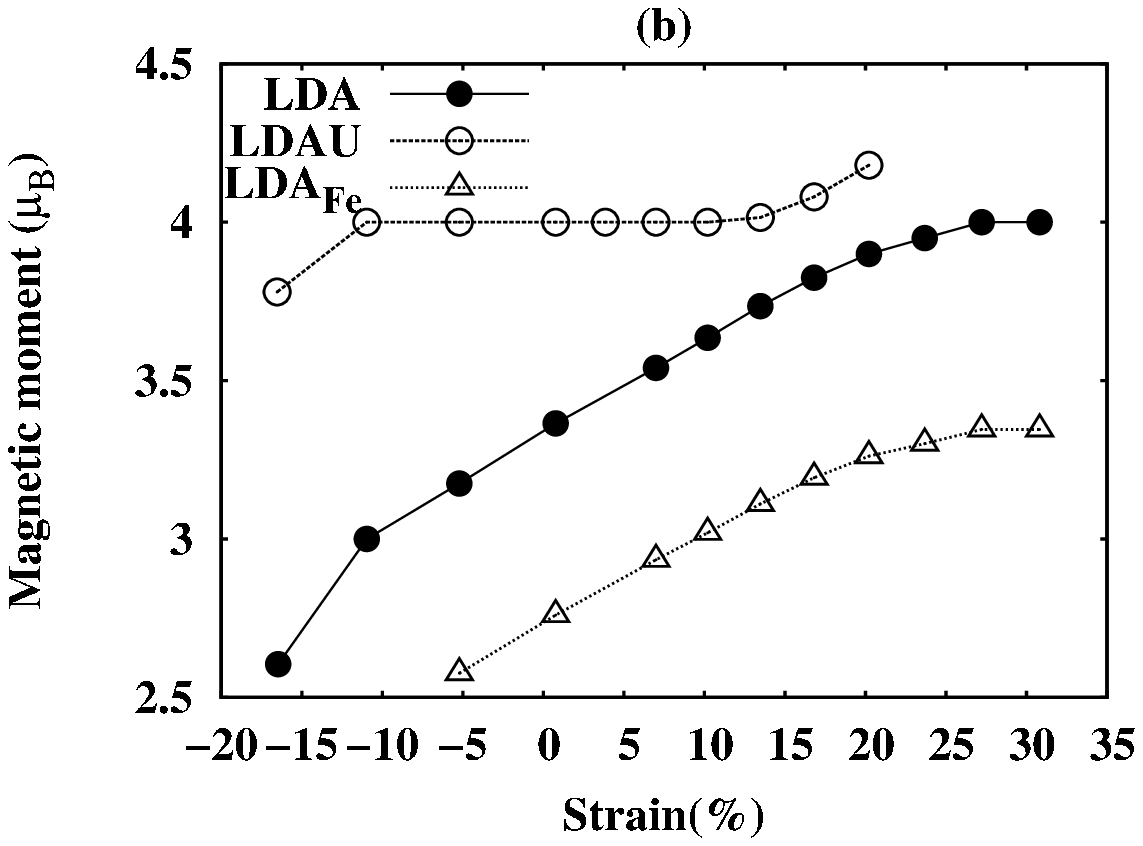}\hspace{0.4cm}\\
\includegraphics [width=0.14\textwidth]{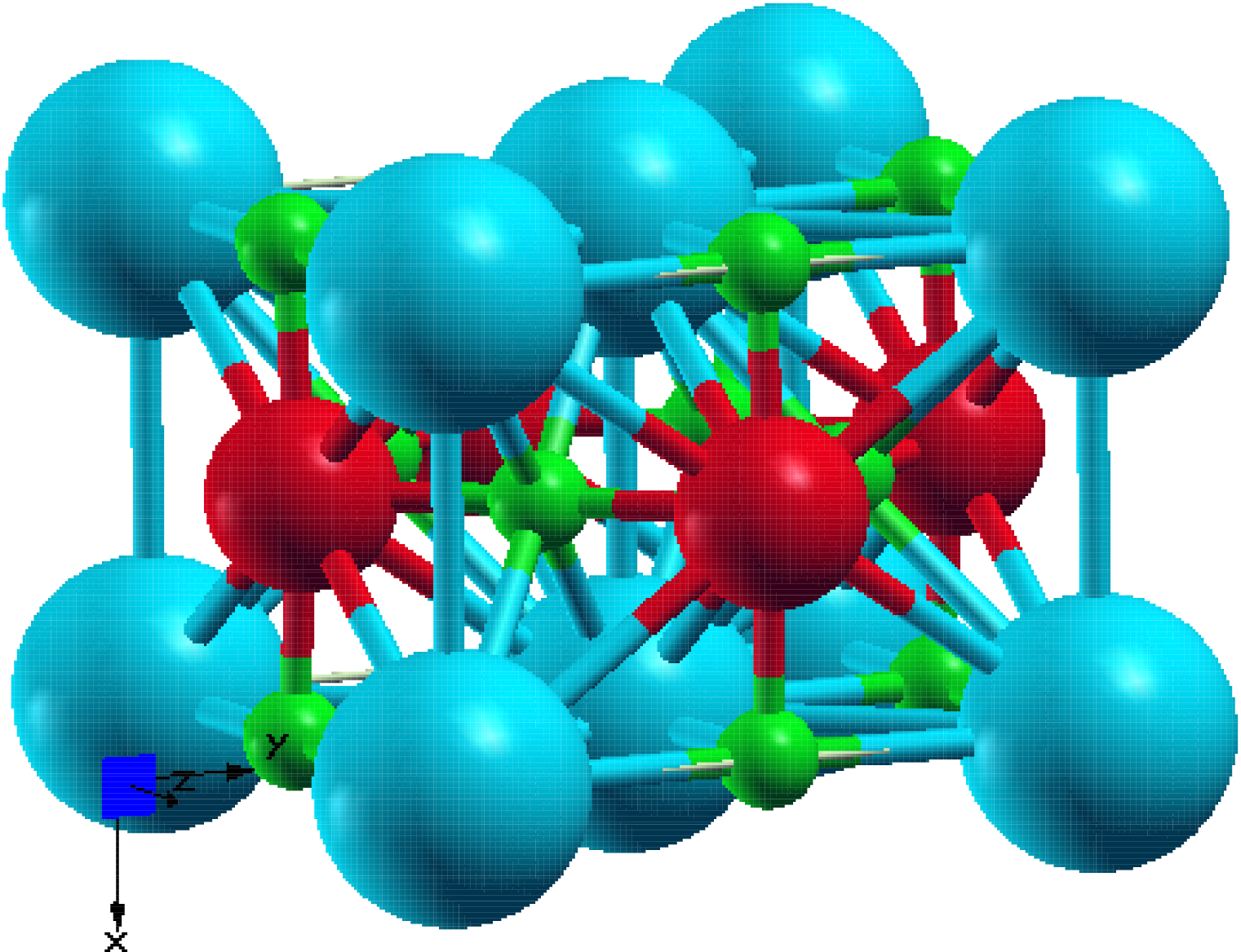} 

\caption{(a) The LSDA calculated total energy/fu vs strain for FM(filled circles) and NM(empty circles) BFO. (b) The calculated total magneitc moment/fu in LSDA(filled circles), LSDA$+U$(empty circles), and the local magnetic moment of Fe in LSDA(triangles) vs strain of BFO. The magnetic moments of BFO are shown in the units of $\mu_{\rm B}$. The crystal structure of orthorhombic BFO is also shown, where cyan, red, and green balls represent Ba, Fe, and O atoms, respectively.}
\label{TE}
\end{figure}

\label{sec:result}
It is essential to optimize the lattice volume of BFO either using the molecular dynamics or the total energy calculations. We followed the later procedure and calculated the total energies in the nonmagnetic (NM) and ferromagnetic (FM) states. Using the equation of state~\cite{birch}, the equilibrium lattice volume $V_{0}$ is determined, which is $56.28\,$\AA$^{3}$/formula unit (fu) ($a=3.82, b= 5.43, c=5.45$\AA). Figure~\ref{TE}(a) shows the strain, which is $\frac{V-V_{0}}{V_{0}}$, vs energy, where positive (negative) strain corresponds to a tensile (compressive) strain. In the whole range of strain, the FM state remains in the lowest energy state.
Using the optimized lattice volume, the formation energy of BFO is calculated\cite{form}, which is $\sim-3.80$ eV/fu. The negative formation energy illustrates that it may not be difficult to synthesize BFO. To see any possible antiferromagnetic (AFM) interactions  between the Fe atoms, the FM and AFM interactions between the Fe atoms were also considered. We found that the FM state is $~0.130$ eV/fu lowered than the AFM state. {It should be mentioned} that the non-stoichiometric BFO may have other magnetic structures.\cite{RFF} Once we have confirmed that the FM state is the ground state of BFO, then we investigated the magnetic and electronic properties of BFO in the FM state under different tensile and compressive strains.

The calculated magnetic moment/fu is shown in Fig.~\ref{TE}(b). It is evident  that the magnetic moment of the lattice is very sensitive to the strain, and the magnetic moment decreases with compression. The magnetic moment increases monotonically with the tensile strain up to $\sim 25\%$. The increment/decrement of the magnetic moment is mainly due to elongation/compression of the Fe-O bond length which further decreases/increases the hybridization between the Fe-$d$ and O-$p$ orbitals that control the charge transfer between the Fe and O atoms.
Beyond  $\sim 25\%$ tensile strain, saturation in the magnetic moments, $4.0\mu_{\rm{B}}$ /fu, i.e., Fe$^{4+}$ can be seen. The local magnetic moments of Ba, Fe, and O were also analysed, and the increase/decrease in the total magnetic moments is mainly contributed by the local magnetic moments of the Fe atoms, which are also shown in the same figure. Small dependence of the O magnetic moments on strain is also observed. The LSDA calculated total magnetic moment of BFO at the equilibrium volume is $3.34\mu_{\rm{B}}$/fu, whereas the local magnetic moments of Fe and O are $2.74\mu_{\rm{B}}$ and $ 0.19\mu_{\rm{B}}$. The integer (saturated) magnetic moment at a large tensile strain is an indication of half-metallic ferromagnetism in BFO. Therefore, we can expect half-metallic ferromagnetism in the strained BFO, which will be discussed in the following paragraphs. The local magnetic moment of Fe (O) at $\sim 25\%$ tensile strain is $3.35\mu_{\rm{B}}$($0.20\mu_{\rm{B}}$) which further shows that the Fe local magnetic moment is mainly responsible for the integer magnetic moment of BFO. The magnetic interactions between the Fe and O atoms remained ferromagnetic due to positive induced polarization at the O sites. 

We observed that the $\sim 25\%$ strained BFO can show interesting magnetic properties, and such a large strain may be  achieved experimentally by depositing BFO on a large mismatched substrate. However, LSDA usually ignores the strong correlation effect between the Fe atoms. If such a strong correlation effect exists between the Fe atoms, then it is expected that BFO may show integer magnetic moment (half-metallicity) without strain. 
We then repeated the above calculations using the LSDA+U approach, and the calculated magnetic moment at each strain is also shown in Fig.~\ref{TE}(b). The total magnetic moment of BFO at $V_{0}$ is $4.0\,\mu_{\rm{B}}$/fu, whereas the local magnetic moments of Fe and O are $3.73\mu_{\rm{B}}$ and $ 0.09\mu_{\rm{B}}$. One can easily notice that including $U$ has increased (decreased) the magnetic moment of Fe (O). It is also interesting to note that the total magnetic moments are remained robust against the strain, and the magnetic moments are saturated to $4.0\,\mu_{\rm{B}}$/fu in the whole strain region. The integer value of the magnetic moment is an indication of half-metallic magnetism in BFO, as we observed in the strained BFO LSDA calculations. However, a drastic change in the magnetic moment can be seen at a large compressive or tensile strain, where the magnetic moment deviates from the integer value suggesting a possible non half-metallic behavior. Note that for $\sim 25\%$  tensile strained, LSDA showed half-metallic behavior and LSDA$+U$ has destroyed the half-metallicity in BFO {as shown below}. {Hence, there is a strong competition between the strain energy, which localizes the minority electrons (in case of elongation), and Coulomb energy that delocalizes electrons at large strain(in case of compression), and strain energy looses at the cost of half-metallicity.}   
From these observations, it can be inferred that a large strain can destroy the half-metallic nature of BFO even if the effect of $U$ is included, which will also be highlighted in the following paragraphs.

\begin{figure}
\includegraphics [width=0.25\textwidth]{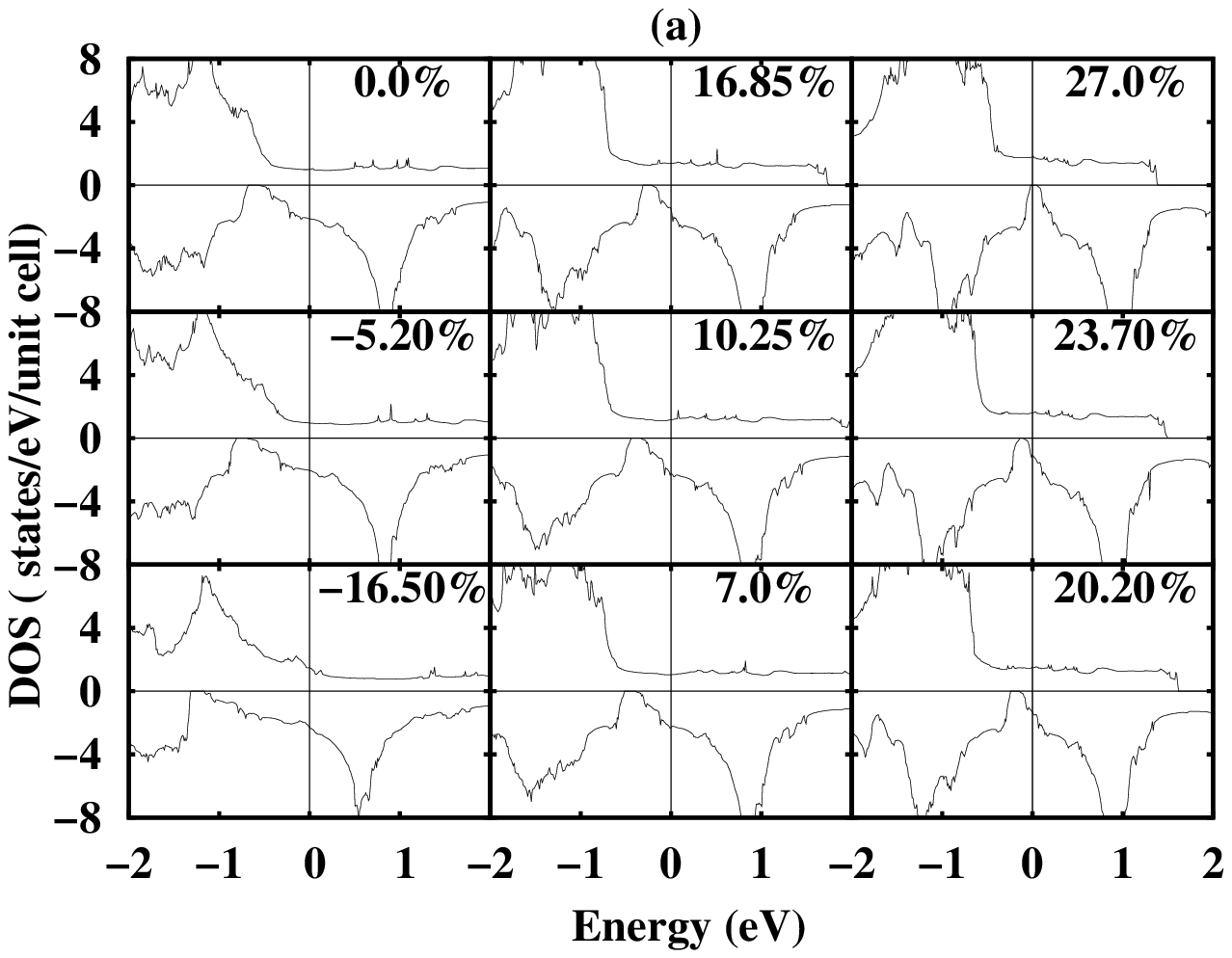}
\includegraphics [width=0.25\textwidth]{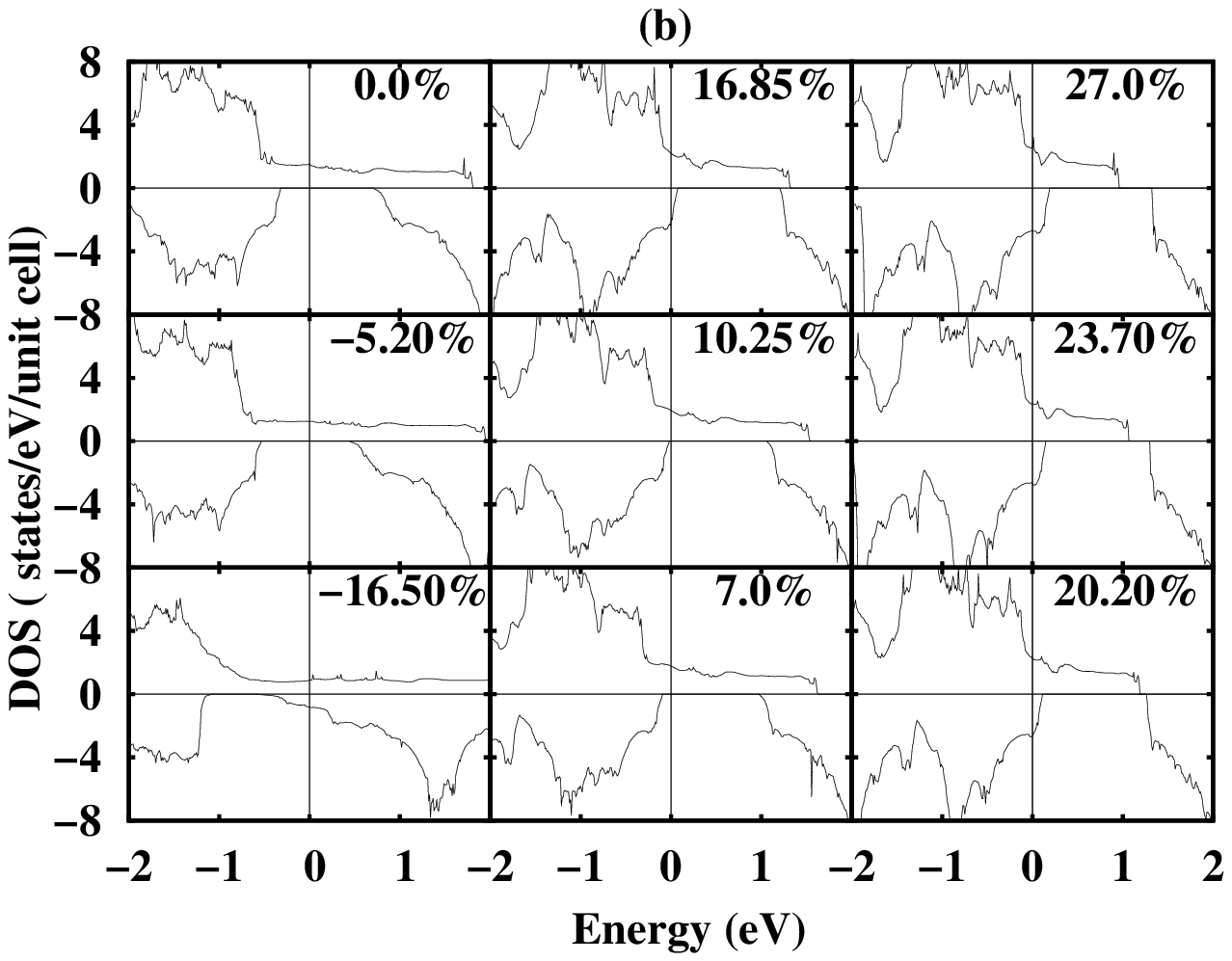}
\caption{(Color online) The calculated total electronic density of states of BFO under different strain in LSDA(a) and LSDA$+U$ (b). All the studied strain are mentioned where $0.0\%$ strain corresponds to the equilibrium structure of BFO. The Fermi energy $E_{\rm F}$ is set to zero. The upper (lower) panels shows majority (minority)-spin DOS.}
\label{T-DOS}
\end{figure}

As shown above that BFO can have integer magnetic moment (expected half-metalic ferromagnetism) at a large strain in the LSDA calculations, however, BFO  shows integer magnetic moment at equilibrium volume and a large strain can suppress the integer magnetic moment (expected non half-metalic ferromagnetism) nature of BFO in the LSDA+U calculations. 
To further confirm these gestures, we calculated the total electronic density of states (DOS) at each strain, and the total DOSs are shown in Fig.~\ref{T-DOS} (a), where $0.0\%$ strain corresponds to the equilibrium volume. At the equilibrium volume, BFO is a  metallic magnetic material with no trace of half-metallicity. The Fermi energy lies in the conduction band, and one can also see a pseudogap in the minority spins just below the Fermi energy. The pseudogap can be shifted to the Fermi energy, if an external strain is imposed.~\cite{gr2010} By a compressive strain, BFO shows metallic behavior and the pseudogap is shifted more below the Fermi energy, i.e, the Fermi energy penetrates more the conduction band.  Therefore, the compressive strain does not help in inducing half-metallicity. {However, on the other hand, for a tensile strain  the Fermi energy shifts towards the pseudogap, and the Fermi energy shifts away from the conduction band, and at a large strain ($~25\%$) the Fermi level lies just at the conduction band edge $E_{c}$ indicating that the minority spin band is completely occupied. Such movement of the Fermi energy in the minority spins state increases the magnetic moment [see Fig.~\ref{TE}(b)].} Beyond $~25\%$ tensile strain, the Fermi energy clearly separates the conduction and valance bands and a bandgap of $\sim 0.07$ eV in the minority spins state can be seen. Such an electronic structure of a FM material is known as a half-metal ferromagnet consistent with the integer magnetic moment as discussed above.

We have shown that strain can easily tune the electronic structure of  BFO. However, for oxides containing transition metals, the strong on-site Coulomb energy and exchange interactions for the $d$ electrons between neighbouring transition metal ions due to the local $d$ orbitals can not be ignored. Knowing that LSDA usually underestimates the size of the bandgap in a material with strongly localized $d$ electrons and even predicts the metallic behvior of meterials that are known to be insulator. We then used LSDA+U to describe the correct electronic structures of BFO under strains, which are shown in Fig.~\ref{T-DOS}(b). The band structure at the optimized volume is represented by $0.0\%$ strain, and one can easily see a large bandgap ($\sim 1.04$ eV) in the minority spins state, and the Fermi level lies in the bandgap--a half-metal ferromagnet. These calculations indicate that the strong Coulomb interaction between the $d$ electrons has overcome the tensile $\sim 25\%$ strain that was required for half-metallic ferromagnetism in BFO in the LSDA calculations. 
The main effect of strain is to shift the Fermi energy towards valence band edge$E_{v}$ in the minority spins state in the strained BFO, and for a large tensile strain $E_{F}$ crosses $E_{v}$. Thus some of the minority spins state becomes unoccupied due to a large tensile strain that further increases the magnetic moment of BFO and no longer remains integer magnetic moment. And the compressive strain shifts the Fermi energy to $E_{c}$ that promotes more electrons from the conduction band to the valance band in the minority spins state and the magnetic moment decreases from its integral value.
Thus a large compressive or tensile strain can destroy the half-metallic nature of BFO when the effect of $U$ is considered.
Therefore, the main effect of $U$ is to push apart the bonding and anti-bonding states that widen the bandgap, but strain shifts the Fermi energy and can drastically change the electronic structure. 
Figure~\ref{T-DOS}(b) also demonstrates that
BFO can retain its half-metallic nature under $\pm \,10\%$ strain, and beyond this strain BFO will have a metallic nature. The half-metallic ferromagnetic behvior is consistent with the integer magnetic moment as discussed in Fig.~\ref{TE}(b) and the decrease/increase in the magnetic moment at a large compressive/tensile strain is due to the non half-metallic property of BFO.

\begin{figure}

\includegraphics [width=.3\textwidth]{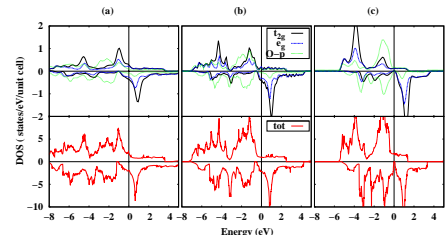}
\caption{(Color online)The LSDA calculated total(lower panel) and partial density of states(upper panel) of BFO under -16.50$\%$ (a), equilibrium (b), and 25.00$\%$ (c) strain. In the upper panel, the solid(black), dashed(blue), and dotted(green) lines shows Fe-$t_{2g}$, Fe-$e_g$, and O-$p$ states, respectively. The Fermi energy $E_{\rm F}$ is set to zero. The upper (lower) panels shows majority (minority)-spin DOS.}
\label{ST-DOSLDA}
\end{figure}

\begin{figure}

\includegraphics [width=.3\textwidth]{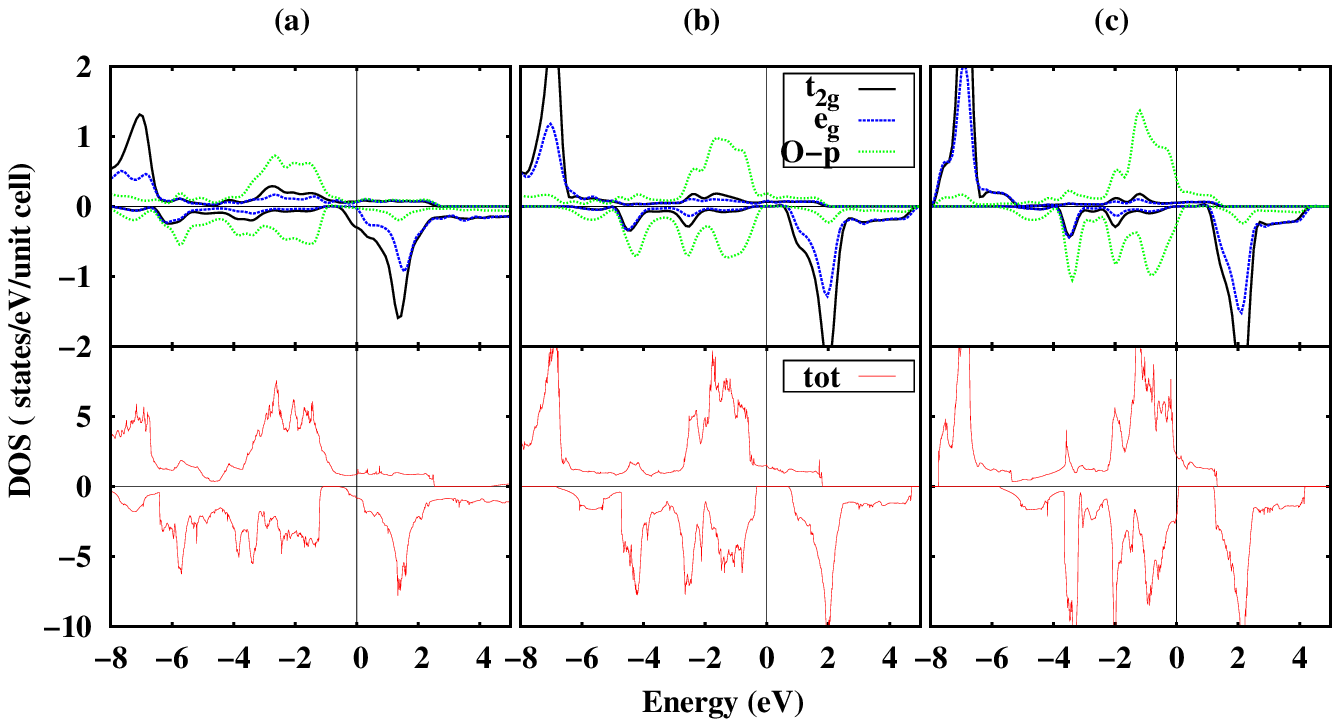}
\caption{(Color online)The LSDA$+U$ calculated total(lower panel) and partial density of states(upper panel) of BFO under -16.50$\%$(a), equilibrium(b), and 16.85$\%$(c) strain.In the upper panel, the solid(black), dashed(blue), and dotted(green) lines shows Fe-$t_{2g}$, Fe-$e_g$, and O-$p$ states, respectively. The Fermi energy $E_{\rm F}$ is set to zero. The upper (lower) panels shows majority (minority)-spin DOS.}
\label{ST-DOSLDAU}
\end{figure}

It is very essential to further investigate the atomic origin of strain or correlation induced half-metallic ferromagnetism in BFO, and why BFO looses its half-metallic nature in the presence of a large strain even when $U$ is included. Figure~\ref{ST-DOSLDA} shows the LSDA total and atom projected orbital resolved DOS of $-16.5\%$ (panel (a)), $0.0\%$ (panel (b)), and $25\%$ (panel (c)) strains.
For $0.0\%$ strain, the partial DOS shows strong $p$-$d$ bonding which gives rise to metallicity in BFO. The Fermi energy crosses the conduction and valence bands, and the minority spins band near the Fermi energy is mainly dominated by the Fe $e_{g}$ and $t_{2g}$ electrons. The degenerate behavior of Fe $e_{g}$ and $t_{2g}$ indicates that the crystal field is not strong enough to remove the degeneracy of the $d$ orbitals.  The tensile strain mainly narrows the $d$ band, and the bandwidth of the $d$ band is also decreased. Due to strong overlap of the Fe-$d$  and O-$p$-orbitals, the O-PDOS also becomes localized and participates in the half-metallicity of BFO. For the compressive strain, the bands become very wide, and these are basically the minority $t_{2g}$ and $e_{g}$ bands that help in producing/destroying the half-metallicity in BFO. The LSDA$+U$ calculated PDOS of$-16.5\%$(panel (b)), $0.0\%$ (panel (b)), and $16.85\%$(panel (c)) strains are also analysed in Fig.~\ref{ST-DOSLDAU}. One can clearly see that at the equilibrium volume, the half-metallicity is mainly caused by the shift of Fe $t_{2g}$ and $e_{g}$ electrons, and these bands are completely occupied in the majority spins state, which were partially occupied without considering $U$. In the minority spins state,  the $t_{2g}$ and $e_{g}$ electrons of Fe are almost unoccupied, and one can see a large bandgap of $1.04$eV due to the unoccupation of these bands. As discussed above that for the compressive strains, the Fermi energy goes deeper into the conduction band, and now it is clear from Fig.~\ref{ST-DOSLDAU}(a) that the half-metallicity is mainly destroyed by $t_{2g}$ and $e_{g}$ bands, and predominantly by the $t_{2g}$ electrons. However, for the tensile strain, the half-metallicity is not destroyed by the $t_{2g}$ and $e_{g}$ electrons of Fe, but by the $p$ electrons of O. The tensile strain shifts the Fermi energy towards $E_{c}$ and the top of the valance band in the minority spins is dominated by the $p$-orbitals of O and the Fe $d$ band is far from the Fermi energy due to the on-site Coulomb repulsion(see Fig.~\ref{ST-DOSLDAU}(c)). Finally, comparing the PDOS of LSDA with LSDA$+U$, it is noticed that the hybridization between Fe-$d$ and O-$p$ electrons is weaker in the LSDA$+U$ case, and the top of the valance band in both the spins state is mainly dominated by the $p$ electrons of oxygen. {Before summarizing our theoretical calculations, we would like to comment on the helical magnetic order, which is observed in cubic BFO.\cite{LiPRB2012} Magnetism is considered to be a cooperative phenomenon and we mainly focused on the collinear behaviour of spin magnetic moments of Fe in orthorhombic BFO. Investigating the helical magnetic order is beyond the scope of the present work.} 

{To summarize, we predicted strain and correlation induced half-metallic ferromagnetism in BFO using DFT calculations. A large strain would be needed for half-metallicity, but no strain is required to have half-metallic ferromagnetism, when the correlation effect between the Fe $d$ electrons is considered. The half-metallic behavior in the strained BFO is mainly caused by the shift of the $t_{2g}$ and $e_{g}$ electrons of Fe from the Fermi energy. Large compressive and tensile strain in the presence of $U$ can destroy the half-metallicity in BFO. The presence (absence) of half-metallic ferromagnetism in BFO is consistent with integer(non-integer) magnetic moment of BFO. Therefore, from these DFT calculations, we can infer that $U$ is playing a vital role in inducing half-metallic ferromagnetism in BFO. Generally, it may be difficult to impose a large strain due to elastic modulus of a material and such a large strain may induce dynamic instability. Therefore, for materials with inherent correlation effect between the transition metals, e.g. Fe,  can show interesting properties (half-metallic magnetism) with no external strain. Strain in a strongly correlated system can be used as a parameter to see the maximum absorption  of the strain to retain its half-metallic behavior, and we found that BFO can withstand $\sim \pm 10\%$ strain before converting into metallic BFO.
As $U$ can be considered as a source to overcome the large tensile strain, therefore, we believe that those materials that can have strong Coulomb interactions are good candidates for applications in the area of spin-electronics or multiferroics. Further experimental work is needed to confirm our predictions.}



\begin{thebibliography}{100}
\bibitem{form} The formation energy $E_{f}$ of BaFeO$_{3}$ was calculated as $E_{f}$=E(BaFeO$_{3}$)-[E(Ba)+E(Fe)+$\frac{3}{2}$E(O$_{2}$)], where E(BaFeO$_{3}$) is the total energy in of orthorhombic BaFeO$_{3}$, E(Ba), E(Fe)
and E(O$_{2}$) are the total energies of BCC Ba, BCC Fe, and oxygen molecule, respectively
\end{thebibliography}
\end{document}